\documentclass{article}[12pt]

\usepackage{amsmath}
\usepackage{amsfonts}

\usepackage{graphicx}
\usepackage{natbib}
\usepackage{theorem}
\usepackage{setspace}
\usepackage{amssymb}
\usepackage{color}
\usepackage{lineno}
\newtheorem{theorem}{Theorem}[section]
\newtheorem{proof}{Proof}[theorem]

\newcommand{\V}{{\mathsf{FP}}}
\newcommand{\T}{{\mathsf{TP}}}
\newcommand{\E}{{\mathbb{E}}}
\newcommand{\Pro}{{\mathbb{P}}}
\newcommand{\pro}{{\mathbb{P}}}
\newcommand{\prob}[1]{\mathbb{P}\paren{#1}}
\newcommand{\Rj}{{\mathcal{R}}}

\newcommand{\paren}[1]{\left(#1\right)}

\newcommand{\sfdp}{\mbox{SFDP}}
\newcommand{\sev}{\mbox{SEV}}

\newcommand{\pvalues}{p_1,\ldots,p_m}

\newcommand{\distsu}{\mathcal{D}}

\newcommand{\Sti}[2]{\genfrac{\{}{\}}{0pt}{}{#1}{#2}}
\renewcommand{\l}{\ell}
\newcommand{\SU}{\mbox{SU}}

\newcommand{\mtc}{\mathcal}

\newcommand{\Pow}{\mbox{Pow}}
\newcommand{\cf}{\mtc{F}}
\newcommand{\lla}{\mtc{L}_{\lambda}}

\begin{document}
\title{Optimality in multiple comparison procedures}
\author{Djalel Eddine Meskaldji\footnote{Signal Processing Laboratory (LTS5), Ecole Polytechnique F\'{e}d\'{e}rale de Lausanne (EPFL), Lausanne, Switzerland. Email: djalel.meskaldji@epfl.ch.} \footnote{This work was supported in part by the FNS grant N$^0$144467.}
\and Jean-Philippe Thiran* \and Stephan Morgenthaler\footnote{FSB/MATHA, Ecole Polytechnique F\'{e}d\'{e}rale de Lausanne (EPFL), Lausanne, Switzerland.}}
\maketitle
\doublespacing
\begin{abstract}
When many (m) null hypotheses are tested with a single dataset, the
control of the number of false rejections is often the principal
consideration. Two popular controlling rates are the probability of making at least one false discovery (FWER) and the expected fraction of false discoveries among  all rejections (FDR). 
Scaled multiple comparison error rates form a new family that bridges the gap between these two extremes. For example, the Scaled Expected Value (SEV) limits the number of false positives relative to an arbitrary increasing function of the number of rejections, that is, $\E(\V/s(R)\vee 1)$. We discuss the problem of how to choose in practice which procedure to use, with elements of
an optimality theory, by considering the number of false rejections $\V$ separately from the number of correct rejections $\T$. Using this framework we will show how to choose an element in the new family mentioned above.
\end{abstract}


\section{Introduction}
The theory of multiple testing is dominated by discussions of error rates and
the procedures that control those rates. The outcome of $m$ tests can be
summarized by the number of true rejections $\T$ (the rejections among the
$m_1$ true alternatives) and the false rejections $\V$ (the rejections among
the $m_0$ true null hypotheses). The total number of rejections is
$R=\T+\V$.\\

With this paper, we want to broaden the discussion to include the optimal
choice of error rate.  This choice depends on the number of tests $m$, the likely size of the alternative effects and the fraction of true nulls $m_0$ among the $m$ null hypotheses. To illustrate why this is so,
consider the following example. If the true alternatives are sparse (small
$m_1$), then the FDR will almost always be better than the FWER, because it
has a better chance of detecting the true alternatives, and yet will not make
many false discoveries. Another situation is when the effect sizes that define
the alternatives are huge, then the FWER is slightly better, because it will
also detect the true alternatives, but will make even fewer mistaken rejections.
As $m_1$ increases, the choice of the FDR becomes problematic due to the definition of the control.
Even a small percentage of a large number of rejections can be sizable.\\

In the aim of bridging the gap between the two extremes, \cite{MeskCER2011} introduced the scaled error rates. The number of false positives is considered with the number of rejections via a scaling function, that is, the ratio $\V/s(R\vee 1)$ is considered and is called the Scaled False discovery Proportion SFDP. \cite{MeskCER2011} derived as well, procedures that control either the quantiles
or the expectation of the SFDP. 
The expectation of the SFDP is called the Scaled Expected Value (SEV) defined by $$\sev_s=\E\left[
\frac{\V}{s(R \vee 1)}\right],$$ where $s(\cdot )$ is a non-decreasing positive function called the scaling function. The Per Family Error rate $\E(\V),$ and the FDR are met by setting $s(R)\equiv1$ and $s(R)=R$ respectively.\\

The procedure that control the SEV under dependence and positive dependence is a step up (SU) procedure that uses the sequence of thresholds $\mathcal {T}_s =(t_i =\frac{s(i)}{m}\alpha)_{1\leq i \leq m}$. This is a scaled version of the LSU procedure proposed by \cite{BenjaminiFDR1995} to control the FDR. This procedure generalizes many multiple comparison procedures. The Bonferroni procedure and the LSU procedure are met by setting $s(i)\equiv1$ and $s(i)= i,\, \forall i \in I,$ respectively. Note that the Bonferroni procedure controls the PFER which implies the control of the FWER by Markov's inequality.\\

%
The choices offered by the scaled error rates opens the question of how to
proceed in practice. We will investigate some aspects of this question in this paper.
%
%
Among the Multiple Comparison Procedures (MCPs), the ones that reject a
maximal number of hypotheses are preferred. This is the extent to which
optimality is investigated. First, we have to find a common optimality
criterion to compare the different error metrics and control procedures. We
propose to measure the worth of each true discovery by the value 1 and the
loss due to a false discovery by $-\lambda$.\\

We present the optimality criterion and discuss the choice of the parameter $\lambda$ in Section~\ref{PartI: section4.2}. In Section~\ref{PartI: section4.3}, we derive asymptotic results for the SEV and we investigate in more details a particular case of scaling functions which is $s(i)=i^\gamma,$ with $\gamma \in [0,1].$ We present different simulations for this particular case in Section~\ref{PartI: section4.4}. Finally, we derive exact calculations for the SFDP under the unconditional mixture effect model using the SU procedure described above. The results are based on Theorem 3.1 of \cite{Roquain2011exact} and obtained immediately when inserting the scaling function at the right places.

\section{Optimality of MCPs}\label{PartI: section4.2}
The general goal of any multiple testing procedure, consists in making $\T$
large while keeping $\V$ small. The two types of rejections are opposites of
each other, but asymmetrical opposites. The prevailing approach consists in deciding on a level and type of control against false rejections (errors of type I) and subject to this constraint to maximize the number of rejections. This is analogous to the Neyman-Pearson approach of bounding the probability of a false rejection and then, given this constraint, maximizing the power. But since there is no agreement on the choice of control in multiple testing, the analogy is not convincing. This approach does not allow one to compare across a spectrum of type I error metrics. Controlling the false discovery rate, for example, can potentially lead to many rejections and is in this sense powerful, but how should this be compared to a method that controls the probability of making at least one erroneous rejection? \\
\subsection{Common optimality criterion}
One may think of the underlying problem in terms of costs. Each true rejection is worth one unit, while each
false rejections leads to a loss of $\lambda \geq 1$. The cost $\lambda$
of a false discovery is a tuning constant to be set by the user. It acts as a
penalty against false discoveries. If $\lambda=1$, the true and the false
discoveries are of equal value, in which case maximizing the gain $R-2\V$ is
equivalent to minimizing $m_1-R+2\V$, the sum of false rejections and false
discoveries. The cost $\lambda$ can also be seen as a shadow price, that is, the value of a
Lagrange multiplier at the optimum. This interpretation appears if we
optimize the number of true rejections under constraints involving the false
discoveries.\\

Based on this loss, the best choice of error rate minimizes the loss function
\begin{equation}
\lla=\lambda \E[\V]-\E[\T]=(\lambda+1)\E[\V]-\E[R].
\end{equation}
with $m_0\geq 1$ and $m_1 \geq 1$.\\

This approach will be unfamiliar to statisticians, who are used to maximizing power under control of the false
rejections. Our criterion allows a mixture of different error rates and
will pick the one best adapted to $\lambda$.
\subsection{Choice of the cost $\lambda$}
Before starting the main question of the paper we give some thoughts about the choice of the price $\lambda$. In the philosophy of multiple testing, $\lambda\geq1$, because the subsequent investigation of any discovery is expensive and being on the wrong track is a grave mistake. In a more refined theory, the cost $\lambda$ should probably rather be seen as a marginal price, which increases with the number of false discoveries, but we will stay with the simpler model of a fixed price per false rejection.\\

To gain further insight, consider a model case, where $m=2$ with $m_0=m_1=1$
and we observe independent test statistics $X_0\sim{\cal
  N}(0,1)$, a unit Gaussian, and $X_1\sim {\cal N}(\Delta,1)$. We are
testing a zero mean vs. a positive mean and the two tests reject if the
observed value exceeds a critical value $\text{cv}>0$. If we reject based on
$X_0$ we have a false rejection and if we reject based on $X_1$ we have a
true rejection. In this case, $\T$ and $\V$ are independent Bernoulli
variables with success probabilities $p_0=1-\Phi(\text{cv})=\Phi(-\text{cv})$
and $p_1=1-\Phi(\text{cv}-\Delta)=\Phi(\Delta-\text{cv})$. The criterion thus
has value $$\lla=\lambda \E[\V] - \E[\T]=\lambda p_0 - p_1\,.$$
For a fixed price $\lambda$, the largest value of the criterion, the optimal gain, is achieved
for the critical value that satisfies $$-\varphi(\Delta-\text{cv}_\text{opt})
+ \lambda \varphi(-\text{cv}_\text{opt})=0,$$ which leads to
$$\text{cv}_\text{opt}=\log(\lambda)/\Delta + \Delta/2.$$
The optimal gain is always positive, increases with $\Delta$ and decreases with $\lambda$. In
this simple model, the two tests are determined by the critical value.\\

For a fixed price $\lambda$, the optimal critical value $\log(\lambda)/\Delta +
\Delta/2$ as a function of the effect $\Delta$ is convex and has a minimum at
$\Delta = \text{cv}_\text{opt}=\sqrt{2\log(\lambda)}$. This is the optimal
test with the minimal level.\\

When $p_0$ is fixed ($p_0=\alpha$), the price paid for a false positive is
\begin{equation}
\lambda(\Delta)=exp\left\{\Delta\left( \Phi^{-1}(1-\alpha)-\frac{\Delta}{2}\right)\right\}.\label{LambdaDeltaEq}
\end{equation}
\begin{figure}[h!]
\centering
\includegraphics[angle=270,width=0.8\textwidth]{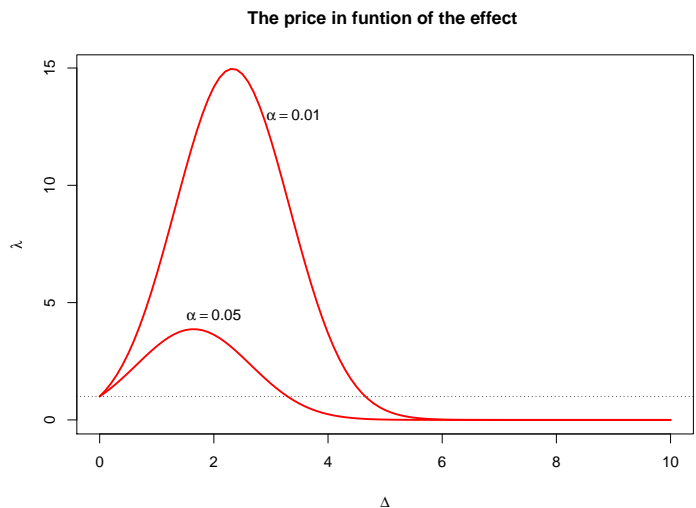}
\caption[The price $\lambda$ in function of the effect $\Delta$.]{The price $\lambda$ in function of the effect $\Delta$ for two values of $\alpha=0.01$ and $0.05$. The curves above $\lambda=1$ are symmetric around $\Phi^{-1}(1-\alpha)$. \label{PartI: PlotLambdaDelta}}
\end{figure}

Equation (\ref{LambdaDeltaEq}) shows that the maximum price that has to be paid corresponds to a situation where the mean of the alternative distribution $\Delta$ is equal
to the critical values of the rejection area. When $\Delta$ becomes small, the
mixture of the observations will more resemble the null distribution and the
probability of rejection decreases until $\alpha$. On the other hand, when
$\Delta$ increases, the probability of detection increases to the point where we
can increase the critical value. When the value of $\Delta$ reaches
$2\Phi^{-1}(1-\alpha)$ the probability of a false negative and false positive
become equal. In this case, $\lambda=1$, which corresponds to the
classification criterion and the critical threshold becomes $\Delta/2$. Figure \ref{PartI: PlotLambdaDelta} shows the behavior of the price $\lambda$ in function of the effect $\Delta$ for two common values of
$\alpha$ namely $\alpha=0.05$ and $\alpha=0.01$. \\

To link this with the classical testing theory, consider the Bonferroni procedure for two one-sided tests with overall FWER of $\alpha$. For example, if $\alpha=0.05$ then $\lambda=3.868132$ and if $\alpha=0.01$ then $\lambda=14.96849$. This gives an idea on the price used in this case. At the very least, this model suggests that the price of a false discovery has to be substantially higher than 1. There has to be a real penalty associated with a
false discovery.
\section{Asymptotically optimal procedure}\label{PartI: section4.3}
\subsection{General results}
For independent tests we can think of the p-values as a mixture of $m_0$
random draws from the uniform distribution and $m_1$ random draws from the
alternative distribution, which might itself be a mixture
distribution. Suppose that $\cf$ is the common distribution of the p-values under
the alternative hypothesis. \cite{Genovese2002operating} showed that
asymptotically (i.e. for large $m$), the LSU procedure corresponds to rejecting the null
hypothesis when the corresponding p-value is less than a threshold $u^*$ where
$u^*$ is the solution of the equation $F(u)=\eta u$ with
 \[
 \eta=\frac{1/\alpha -\pi_0}{1-\pi_0},
 \]
where $F$ is the cumulative probability distribution of $\cf$, and $\pi_0 = m_0/m.$ They showed also that the LSU procedure is intermediate between the Bonferroni procedure (corresponding to $\alpha/m$) and non-multiplicity correction (corresponds
to $\alpha$). Clearly, this shows that the gain in power of the LSU
procedure is due to an increase of the expected number of false positives
from $\pi_0 \alpha/m$ to $\pi_0 U^*$. We give in this section similar results for the SEV.\\

Suppose that the scaling function $s$ is such that
$$\E\paren{\frac{\V}{s(R)}}=\frac{\E(\V)}{\E\paren{s(R)}}+\xi(m),$$
where $\xi(m)\rightarrow 0$ when $m \rightarrow \infty.$ In this case, $u^*$ satisfies
$$
\frac{m_0 u}{s\left(m_{0}u+\left( m-m_{0}\right) F\left(u\right)\right)}%
=\alpha.
$$
Under certain assumptions on $s$, $u^{\ast }$ is the unique solution of
\begin{equation}
s^{-1}\left( \frac{u m_0}{\alpha }\right)=m_{0}u+\left( m-m_{0}\right)
F\left( u\right),
\end{equation}
which leads to
\begin{equation}
\left( m-m_{0}\right) F\left( u^{\ast }\right)  =s^{-1}\left( \frac{%
u^{\ast }m_0}{\alpha }\right) -m_{0}u^{\ast}.\label{PartI: Equ. findUstar}
\end{equation}
The optimization criterion $\mathcal{L}_{\lambda}$ becomes
\begin{eqnarray}
\nonumber \mathcal{L}_{\lambda} &\simeq &\lambda m_{0}u^{\ast }- \left( m-m_{0}\right) F\left( u^{\ast
}\right) \\
\nonumber &=&\lambda
m_{0}u^{\ast }-s^{-1}\left( \frac{u^{\ast }m_0}{\alpha }\right) -m_{0}u^{\ast } \\
\nonumber &=&\left( \lambda -1\right)\alpha
\left(  m_{0}/alpha\right) \left( \frac{u^{\ast }}{\alpha }\right)-s^{-1}\left( \frac{u^{\ast }m_0}{\alpha }\right)  \\
&=&\left( \lambda -1\right)  \alpha v-s^{-1}\left( v\right),\label{PartI: Equ. LossFunctionAssymp}
\end{eqnarray}%
where $v=\frac{u^{\ast }m_0}{\alpha }.$ \\
\subsection{A particular case}
Consider now, the particular case of $s(R)=R^\gamma,$ with $\gamma\in[0,1]$. Then, the SEV becomes
$\E\left( \frac{\V}{R^\gamma}\right).$ This family of error rates
includes the PFER and the FDR for $\gamma=0$ and $1$ respectively.
\cite{MeskCER2011} showed that the family of thresholds
$t_i=\alpha s_\gamma(i)/m=\alpha i^\gamma/m$ provides weak control of the
FWER at a common level $\alpha$. This defines the family of MCPs we will
consider. They are indexed by the parameter $0\leq\gamma\leq 1$ and will be
denoted by SU$_\gamma$. When $\gamma=0$, the Bonferroni procedure results,
while $\gamma=1$ corresponds to the LSU procedure.\\

The $\sev$ for this family, can be approximated by
\[
\E\left( \frac{\V}{R^\gamma}\right) =\frac{m_{0}p_{0}}{\left( m_{0}p_{0}+m_{1}p_{1}\right) ^{\gamma }}+\mathcal{O}(m^{-\gamma/2}).
\]
\begin{proof}
Set $$g(\V,\T) =\frac{\V}{s(\V+\T)}\,.$$
Then, we have
$$\frac{\partial g(\V,\T)}{\partial \V} =\frac{(1-\gamma)\V+\T}{(\V+\T)^{\gamma+1}},$$
and
$$\frac{\partial g(\V,\T)}{\partial \T} =-\frac{\gamma \V}{\left(\V+\T\right)^{\gamma + 1}}\,.$$
Let $p_0$ and $p_1$ be the probabilities of having a false positive and a true positive respectively. Let also, $\mu_{\V}$ and $\mu_{\T}$ be the expectations of $\V$ and $\T$ respectively.\\

We have $\mu_{\V}=m_0 p_0$ and $\mu_{\V}=m_1 p_1$ under the independence assumption. We use the delta method to provide an approximation for $\E\left( \frac{\V}{R^\gamma}\right)$.
\[
E\left( \frac{\V}{s(R)}\right) \approx \frac{\mu _{\V}}{(\mu _{\V+\T})^{\gamma}}=\frac{%
m_{0}p_{0}}{\left( m_{0}p_{0}+m_{1}p_{1}\right) ^{\gamma }}
\]%
and
\[
Var\left( \frac{\V}{s(\V+\T)}\right) \approx \left(\partial_{\V} g(\mu_{\V},\mu_{\T})\right)^2 Var(\V)+ \left(\partial_{\T} g(\mu_{\V},\mu_{\T})\right)^2 Var(\T)
\]
since $Cov(\V,\T)=0$ by independence.\\

For $s(R)=R^\gamma\,,$ the variance becomes
\begin{eqnarray*}
Var\left( \frac{\V}{R^\gamma}\right) &\approx& \left(\frac{(1-\gamma)\mu_{\V}+\mu_{\T}}{(\mu_{\V}+\mu_{\T})^{\gamma+1}}\right)^2 m_{0}p_{0}(1-p_{0})+
 \left(\frac{\gamma \mu_{\V}}{\left(\mu_{\V}+\mu_{\T}\right)^{\gamma + 1}}\right)^2 m_{1}p_{1}(1-p_{1})\\
  &=& \frac{m_0 p_0}{{(\mu_{\V}+\mu_{\T})^{2\gamma+2}}}\left[\left((1-\gamma)\mu_{\V}+\mu_{\T}\right)^2 (1-p_{0})+
 \left(\gamma^2 \mu_{\V}\right) m_{1}p_{1}(1-p_{1})\right].
\end{eqnarray*}
We have,
\[\mu_\V=m_0 p_0\leq m\gamma,
\]
\[
\left((1-\gamma)\mu_{\V}+\mu_{\T}\right)^2 (1-p_{0})\leq((1-\gamma)m^\gamma +m)^2=C_1 m^2,
\]
\[
\left(\gamma^2 \mu_{\V}\right) m_{1}p_{1}(1-p_{1})\leq \gamma^2 m^\gamma m=C_2 m^{\gamma +1},
\]
and
\[
(\mu_{\V}+\mu_{\T})^{2\gamma+2}\geq C m^{2\gamma+2},
\]
where $C_1$, $C_2$ and $C$ are constants. This leads to,
\[
Var\left( \frac{\V}{R^\gamma}\right)\leq m\gamma \frac{C_1 m^2+C_2 m^{\gamma +1}}{C m^{2\gamma+2}}=\mathcal{O}(m^{-\gamma}).
\]
Hence,
\[
E\left( \frac{\V}{R^\gamma}\right) =\frac{m_{0}p_{0}}{\left( m_{0}p_{0}+m_{1}p_{1}\right) ^{\gamma }}+\mathcal{O}(m^{-\gamma/2}).
\]
\end{proof}

When applying SU$_\gamma$, Equation (\ref{PartI: Equ. findUstar}) becomes
\[
\left( m-m_{0}\right) F\left( u^{\ast }\right)  =\left( \frac{%
u^{\ast }m_0}{\alpha }\right)^{\frac{1}{\gamma}} -m_{0}u^{\ast},
\]
and the expected loss of (\ref{PartI: Equ. LossFunctionAssymp}) becomes
$$\lla=\lambda \E[\V_\gamma] -\E[\T_\gamma] = \left( \lambda -1\right)  \alpha v - v^{\frac{1}{\gamma }} \,.$$
The loss $\lla$ is minimized when
\[
\frac{\partial \mathcal{L}}{\partial \mathcal{\gamma }}=\frac{\partial v}{%
\partial \mathcal{\gamma }}\cdot \left[ -\frac{\log v}{\mathcal{\gamma }^{2}}%
\cdot v^{\frac{1}{\gamma }}-\left( \lambda -1\right) \alpha
 \right]=0,
\]
which implies that
\[
\Rightarrow -\frac{\log v}{\mathcal{\gamma }^{2}}\cdot v^{\frac{1}{\gamma }%
}=\left( \lambda -1\right) \alpha.
\]

Finally, the asymptotically optimal value of $\gamma$ for a given unit price $\lambda$ is obtained by solving the system:
\begin{equation}
\begin{array}{l}
\left( m-m_{0}\right) F\left( u^{\ast }\right)  =\left( \frac{%
u^{\ast }m_0}{\alpha }\right)^{\frac{1}{\gamma}} -m_{0}u^{\ast},\label{gamma1}\\
-\frac{\log \left( u^{\ast }m_0/\alpha\right)}{\gamma ^{2}}\cdot \left( u^{\ast }m_0/\alpha\right)^{\frac{1}{\gamma}%
}=\left( \lambda -1\right)\alpha.
\end{array}
\end{equation}

\section{Simulations}\label{PartI: section4.4}
A simple choice for $F$ is the distribution of the p-value one
obtains from a standardized Gaussian test statistic which under the
alternatives is shifted to the right by a common value $\Delta >0$. The
distribution of the p-values for one-sided tests is then $F_1(u) =
1-\Phi(z_{1-u}-\Delta)$ where $z_u =\Phi^{-1}(u)$. The three parameters
$m_0$, $m_1$ and $\Delta$ characterize a multiple testing problem of the kind
we are going to simulate.\\

\begin{figure}[h]
\includegraphics[angle=270, width=\textwidth]{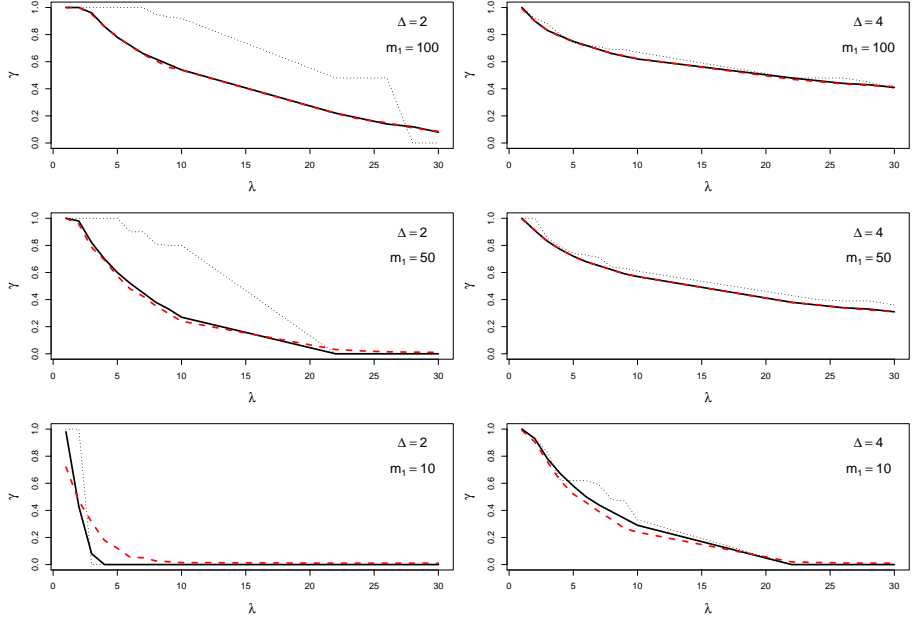}
\caption[The optimal value of $\gamma$ as a function of the penalty $\lambda$ for a false positive when testing $m=1000$ tests.]{The optimal value of $\gamma$ as a function of the penalty $\lambda$ for a false positive when testing $m=1000$ tests, in various situations. In each panel, three curves are plotted. The first curve is obtained by Monte Carlo simulations (points), the second one is obtained by the asymptotic theory assuming that $m_0$ and $\Delta$ are known (solid line) and the third curve is obtained by asymptotic theory with $m_0$ and $\Delta$ estimated by an EM algorithm (dashed). \label{PartI: FigOpt1000}}
\end{figure}

\begin{figure}[h]
\includegraphics[angle=270, width=\textwidth]{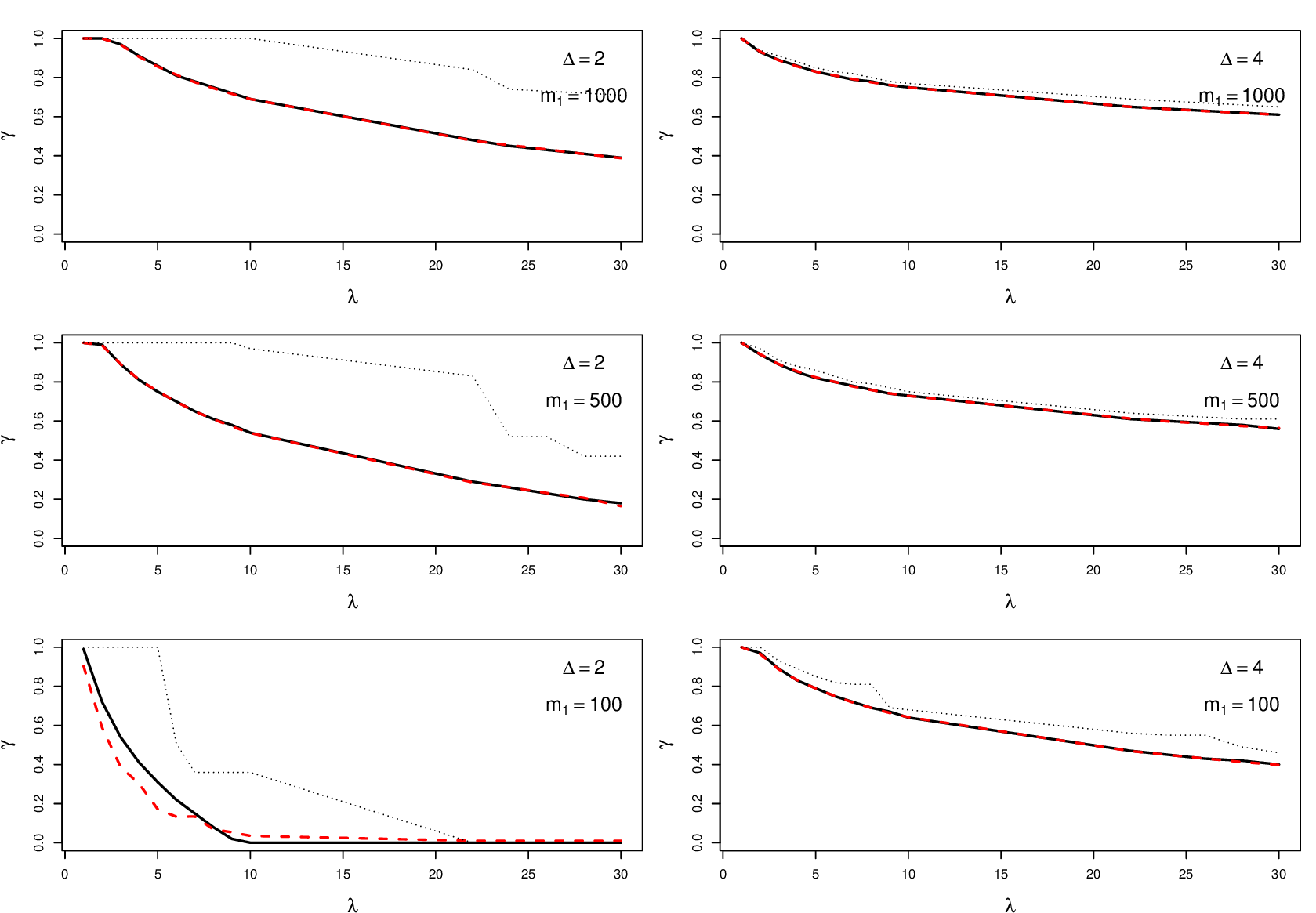}
\caption[The optimal value of $\gamma$ as a function of the penalty $\lambda$ for a false positive when testing $m=10000$ tests.]{The optimal value of $\gamma$ as a function of the penalty $\lambda$ for a false positive when testing $m=10000$ tests, in various situations. In each panel, three curves are plotted. The first curve is obtained by Monte Carlo simulations (points), the second one is obtained by the asymptotic theory assuming that $m_0$ and $\Delta$ are known (solid line) and the third curve is obtained by asymptotic theory with $m_0$ and $\Delta$ estimated by an EM algorithm (dashed).\label{PartI: FigOpt10000}}
\end{figure}

We consider multiple comparisons situations with either $m=1000$ or $m=10000$ tests. We consider $m_1=10, 50$ and $m_1=100$ when $m=1000$, and $m_1=100, 500$ and $m_1=1000$ when $m=10000$. The distribution of the test statistics is the same as in the above model situation with the alternative
effect equal to $\Delta=2$ or $4$. The protection level is $\alpha=0.05$. Figures \ref{PartI: FigOpt1000} and \ref{PartI: FigOpt10000} show the value of $\gamma$ to be used in the case of $s(i)=i^\gamma$ in order to
minimize the expected loss. In each Panel, three curves are plotted. First, the optimal value of $\gamma$ obtained by Monte Carlo simulation. Second, the value obtained by numerically resolving the system of equations (\ref{gamma1}) when the parameters $m_0$ and $\Delta$ are supposed to be known. The third case is identical to the second one except that the two parameters $m_0$ and $\Delta$ are estimated by using the library "mixtools" in the "R" software. The optimal value of $\gamma$ decreases as the penalty $\lambda$ for each false discovery increases. The value $\gamma=1$
which corresponds to the LSU procedure is only optimal for relatively
small penalties, for larger and more reasonable values it quickly drops
towards $\gamma=0.5$ if there are few true alternatives and towards
$\gamma=0.7$ otherwise. For $m=1000$, the effect $\Delta=2$ is relatively
small and hard to detect. For a larger and more easily detectable effect, the
values of $\gamma$ drop even quicker. The value $\gamma=0.5$ is a good
default choice if little is known about the number of alternatives and the
effect size. The optimal value of $\gamma$ obtained asymptotically seems to underestimate the real optimal value, especially, when $\Delta=2$. This underestimation leads to a stricter control of the false positives.
\section{Exact calculations of the SFDP in the SU case under the unconditional independent model}\label{PartI: section4.5}
The aim of this section is to provide exact expressions for the $\kappa$-th moment of the SFDP, the SEV and the power, for any scaling function $s$, when using the SU procedure with thresholds collection $\mathcal {T}_s =(t_r =\frac{s(r)}{m}\alpha)_{1\leq r \leq m}$. The results of the section are based on the work of \cite{Roquain2011exact}, who provided new techniques to derive exact calculations for the FDP and the FDR.\\

Consider the so-called "two-groups mixture model" introduced by \cite{Efron2001} in which $H_i=0$ with probability
$\pi_0$. Let be $G(u)=\pi_0 F_0(u)+(1-\pi_0)F_1(u)$ the common c.d.f. of the p-values, where $F_0$ is the null c.d.f. and $F_1$ is the alternative c.d.f.. This model is called the \emph{unconditional model}. In addition, when the p-values $\pvalues$ are independent, the model is called the \emph{unconditional independent model}.\\

For any $r\geq 0$ and a threshold sequence $\mathcal{T}= (t_1,...,t_r)$, we denote \citep[see][]{Roquain2011exact}
\begin{equation}
\Psi_r(\mathcal{T}) = \Psi_k(t_{1},...,t_{r}) =\prob{U_{(1)}\leq t_{1}, ..., U_{(r)}\leq t_r},\label{equ_psi}
\end{equation}
where $(U_i)_{1\leq i\leq r}$ is a sequence of $r$ random variables i.i.d. uniform on $[0,1]$, with the convention $\Psi_0(\cdot)=1$.

We also introduce the following quantity. For a thresholds sequence $\mathcal{T}=(t_r)_{1\leq r \leq m}$ and $r\geq 0$, $r\leq m$, we define
\begin{align}
\distsu_m(\mathcal{T},r) &=  {m \choose r} (t_r)^r \Psi_{m-r}\big( 1-t_m,...,1-t_{r+1}\big)\label{equ_for_distsu}.
\end{align}

We have that $$\sum_{r=0}^m\distsu_m(\mathcal{T},r)=1$$ for any thresholds sequence $\mathcal{T}$ \citep[see][]{Roquain2011exact}.\\

Recall that the $\kappa$-th moment ($\kappa \geq 1$) of random variable $X$ following a binomial distribution, $X\sim \mathcal{B}(n,p),$ is given by $\E[ X^\kappa]=\sum_{\l=1}^{\kappa \wedge n} \frac{n!}{(n-\l)!} \Sti{\kappa}{\l} p^\l$,
where $\Sti{\kappa}{\l}$ are the Stirling numbers of the second kind, defined by $\Sti{\kappa}{0}=0$, $\Sti{\kappa}{\l}=0$ for $\l>\kappa$,  $\Sti{1}{1}=1$ and the recurrence relation, $ \forall 1\leq \l \leq  \kappa+1$,
$$
\Sti{\kappa+1}{\l} = \l \Sti{\kappa}{\l} + \Sti{\kappa}{\l-1}.
$$
The following theorem is stated and demonstrated by \cite{Roquain2011exact}.
\begin{theorem}\label{main_indep}
When testing $m\geq 2$ hypotheses, consider a SU procedure with thresholds sequence $\mathcal{T}$ and rejection set $\Rj(\mtc{T})$. Then for all $\pi_0\in [0,1]$, we have under the unconditional independent model, for any $r\geq 1$,
\begin{align}
|\Rj \cap I_0 |=\V \: \mbox{ given }\: R\equiv| \Rj(\mathcal{T}) |=r \:\:\:\:\sim\:\: \mathcal{B}\bigg(r, \frac{\pi_0 F_0(t_r)}{G(t_r)}\bigg).\label{equ_distrFDP}
\end{align}
\end{theorem}
From this theorem, we derive the following formulas.
For any $x\in(0,1)$
\begin{equation}
\Pro[\sfdp \leq x ]= \sum_{r=0}^{m} \sum_{j=0}^{ \lfloor x s(r) \rfloor} {{r} \choose {j}} \bigg(\frac{\pi_0F_0(t_r)}{G(t_r)}\bigg)^{j} \bigg(\frac{\pi_1F_1 (t_r)}{G(t_r)}\bigg)^{r-j}  \distsu_m\big( [G(t_{j})]_{1\leq j\leq m},r\big),
\label{equ_FDP_indep}
\end{equation}
where we used the fact that $\pro (R=r)=\distsu_m\big( [G(t_{j})]_{1\leq j\leq m},r\big)$ \citep[see][]{Roquain2011exact}.

\begin{equation}
\E[\sfdp^{\kappa} ]= \sum_{\l=1}^{\kappa \wedge m} \frac{m!}{(m-\l)!}\Sti{\kappa}{\l} \pi_0^\l \sum_{r=\l}^{m} \frac{F_0(t_r)^\l}{s(r)^\kappa}  \: \distsu_{m-\l}\big(  [G(t_{j+\l})]_{1\leq j\leq m-\l} ,r-\l\big).
\label{equ_mom_FDP_indep}
\end{equation}

\begin{equation}
\sev=\pi_0 m \sum_{r=1}^{m} \frac{F_0(t_r)}{s(r)}  \: \distsu_{m-1}\big(  [G(t_{j+1})]_{1\leq j\leq m-1} ,r-1\big).\label{equ_FDR_indep}
\end{equation}

We can apply \eqref{equ_FDR_indep} in the case where $t_r=\alpha s(r) /m$, to deduce that $\sev=\pi_0\alpha$, in the unconditional model. Furthermore, \cite{Roquain2011exact} derived a formula for the power of any SU procedure with thresholds sequence $\mtc{T}$.
\begin{equation}
\Pow(\SU(\mathcal{T})) =  \sum_{r=1}^{m} F_1( t_r)  \: \distsu_{m-1}\big([G(t_{j+1})]_{1\leq j\leq m-1},r-1\big).
\label{equ_Pow_indep}
\end{equation}

When using the thresholds sequence $\cal T_s$ with $t_r=\alpha s(r) /m$, the power becomes
$$\Pow(T_s)  =  \sum_{r=1}^{m} F_1( \alpha s(r)/m)   {{m-1} \choose {r-1}}  (G(\alpha s(r)/m))^{r-1} \Psi_{m-r}\big( 1-G(\alpha m/m),...,1-G(\alpha({r+1})/m)\big).$$

These formulas can help to provide the optimal choice of the scaling function that maximizes a certain criterion of optimality.
\section{Conclusion}
We discussed in this paper ideas on how to choose a scaling function in multiple comparisons. The framework in which we studied this choice used a new point of view, different from the classical view of level and power. The classical approach needs to be rethought and adapted to the multiple comparisons context with large numbers of hypotheses. Under the proposed framework, we derived asymptotic results, especially for a particular family of scaling functions. In a simulation study we showed that an intermediate choice is usually preferable. We also provided exact formulas for the SFDP and the SEV. These formulas can be used in future investigations of the optimal choice of scaling functions.
\bibliographystyle{apalike}
\bibliography{D:/Thesis/Bib}

\begin{thebibliography}{}

\bibitem[Benjamini and Hochberg, 1995]{BenjaminiFDR1995}
Benjamini, Y. and Hochberg, Y. (1995).
\newblock {Controlling the false discovery rate: a practical and powerful
  approach to multiple testing}.
\newblock {\em J. Roy. Statist. Soc. Ser. B}, 57(1):289--300.

\bibitem[Efron et~al., 2001]{Efron2001}
Efron, B., Tibshirani, R., Storey, J., and Tusher, V. (2001).
\newblock {Empirical Bayes Analysis of a Microarray Experiment}.
\newblock {\em Journal of the American Statistical Association}, 96:1151--1160.

\bibitem[Genovese and Wasserman, 2002]{Genovese2002operating}
Genovese, C. and Wasserman, L. (2002).
\newblock {Operating characteristics and extensions of the false discovery rate
  procedure}.
\newblock {\em Journal of the Royal Statistical Society: Series B (Statistical
  Methodology)}, 64:499--517.

\bibitem[Meskaldji et~al., 2011]{MeskCER2011}
Meskaldji, D.~E., Thiran, J.-P., and Morgenthaler, S. (2011).
\newblock {A comprehensive error rate for multiple testing}.
\newblock {\em ArXiv e-prints}.

\bibitem[Roquain and Villers, 2011]{Roquain2011exact}
Roquain, E. and Villers, F. (2011).
\newblock Exact calculations for false discovery proportion with application to
  least favorable configurations.
\newblock {\em The Annals of Statistics}, 39:584--612.

\end{thebibliography}

\end{document}